\documentclass[letterpaper,12pt]{JHEP3}
\usepackage{amsmath,amssymb}
\usepackage{latexsym}
\usepackage{epsfig}
\usepackage{cite}
\usepackage[english]{babel}

\newcommand{\eq}{\begin{equation}}
\newcommand{\feq}{\end{equation}}
\newcommand{\eqn}{\begin{eqnarray}}
\newcommand{\feqn}{\end{eqnarray}}
\newcommand{\arr}{\begin{eqnarray*}}
\newcommand{\farr}{\end{eqnarray*}}

\font\mybb=msbm10 at 12pt
\def\bb#1{\hbox{\mybb#1}}

\def\bR {\bb{R}}

\def\bC {\bb{C}}

\title{Supersymmetric AdS$_4$ black holes and attractors}

\author{Sergio L.~Cacciatori$^{ac}$ and Dietmar Klemm$^{bc}$ \\
$^a$ Dipartimento di Scienze Fisiche e Matematiche, \\
\hspace*{0.15cm} Universit\`a dell'Insubria, \\
\hspace*{0.15cm} Via Valleggio 11, I-22100 Como. \\
$^b$ Dipartimento di Fisica dell'Universit\`a di Milano, \\
\hspace*{0.15cm} Via Celoria 16, I-20133 Milano. \\
$^c$ INFN, Sezione di Milano, Via Celoria 16, I-20133 Milano. \\
}
\preprint{IFUM-947-FT}

\abstract{Using the general recipe given in arXiv:0804.0009, where all
timelike supersymmetric solutions of ${\cal N}=2$, $D=4$ gauged supergravity
coupled to abelian vector multiplets were classified, we construct
the first examples of genuine supersymmetric black holes in AdS$_4$
with nonconstant scalar fields. This is done for various choices of the
prepotential, amongst others for the STU model. These solutions permit to
study the BPS attractor flow in AdS. We also determine the most general
supersymmetric static near-horizon geometry and obtain the attractor equations
in gauged supergravity. As a general feature we find the presence of flat
directions in the black hole potential, i.e., generically the values
of the moduli on the horizon are not completely specified by the charges.
For one of the considered prepotentials, the resulting moduli space is
determined explicitely. Still, in all cases, we find that the black hole
entropy depends only on the charges, in agreement with the attractor
mechanism.
}


\begin{document}

\section{Introduction}
\label{intro}

Since their discovery, the physics of black holes has raised several fascinating
problems and puzzles, whose resolution is believed to be crucial for the
construction of a future quantum theory of gravity. Indeed, much of what we
presently know on quantum effects in strong gravitational fields comes from
the study of black holes. Of particular interest in this context are black holes
preserving a sufficient amount of supersymmetry, which allows (owing to
non-renormalization theorems) to extrapolate a computation of the entropy at
weak string coupling (when the system is generically described by a configuration
of strings and branes) to the strong-coupling regime, where a description in
terms of a black hole is valid \cite{Strominger:1996sh}. These entropy
calculations have been essential for our current understanding of black hole
microstates.

In this paper, we shall construct the first examples of genuine BPS black holes
in four-dimensional anti-de~Sitter space (AdS$_4$) with nontrivial scalar fields
turned on\footnote{Actually we consider also models with scalar potentials that
have no critical points. This leads to black holes that asymptote to curved
domain walls.}. The theory under consideration is ${\cal N}=2$, $D=4$ gauged
supergravity coupled to abelian vector multiplets, whose timelike
supersymmetric backgrounds were classified in \cite{Cacciatori:2008ek}. The
results of \cite{Cacciatori:2008ek} provide a systematic method to construct BPS
solutions, without the necessity to guess some suitable ansaetze. This
facilitates much our analysis here.

The motivation for our interest in supersymmetric AdS black holes is twofold:
First, since the discovery of ${\cal N}=6$ Chern-Simons-matter
theories \cite{Aharony:2008ug}, which (in a certain limit) are dual to type
IIA string theory on AdS$_4$$\times$$\bC P^3$, there has been much interest in
supersymmetric geometries that asymptote to AdS$_4$. In principle, the
AdS$_4$/CFT$_3$ correspondence should allow to compute the microscopic entropy
of AdS$_4$ black holes and to compare it then with the macroscopic
Bekenstein-Hawking result. This would be very tempting to do for the solutions
that we shall present below.
The second reason is the attractor mechanism \cite{Ferrara:1995ih,Strominger:1996kf,Ferrara:1996dd,Ferrara:1996um,Ferrara:1997tw}, which states that the
scalar fields on the horizon and the entropy are independent of the asymptotic
values of the moduli\footnote{Note that this is believed to be the
explanation \cite{Dabholkar:2006tb} of the fact that the Bekenstein-Hawking
entropy of many extremal black holes coincides with a weak coupling
calculation \cite{Kaplan:1996ev,Horowitz:1996ac,Dabholkar:1997rk,Emparan:2006it},
despite the absence of supersymmetry.}. (The scalars are attracted towards
their purely charge-dependent horizon values). 
Given the importance of the attractor mechanism, it would be very interesting to
study the BPS attractor flow in AdS. Some work in this direction has been done
in \cite{Morales:2006gm,Bellucci:2008cb}\footnote{For an analysis of the
attractor mechanism in ${\cal N}=2$, $D=4$ supergravity with SU$(2)$ gauging
cf.~\cite{Huebscher:2007hj}.}, but these papers consider
non-supersymmetric attractors, since up to now no BPS black holes in AdS$_4$
with nonconstant scalars were known\footnote{The solutions found in
\cite{Sabra:1999ux} are naked singularities.}.

Notice that in gauged supergravity, the moduli fields have a potential, and
typically approach the critical points of this potential asymptotically, where
the solution approaches AdS. Thus, unless there are flat directions in the
scalar potential, the values of the moduli at infinity are completely fixed (in
terms of the gauge coupling constants), and it would thus be more precise to
state the attractor mechanism in AdS in the form: "The entropy is determined
entirely by the charges, and is independent of the values of the moduli on the
horizon that are not fixed by the charges". Indeed, we shall encounter below
several examples where the scalars on the horizon are not completely specified
by the charges, i.e., there are flat directions in the black hole potential and
hence a nontrivial moduli space. Yet, in all the cases considered here, the
entropy depends only on the charges. Unfortunately, we found no way to prove
this in general.

The remainder of this paper is organized as follows: In the next section, we
briefly review ${\cal N}=2$, $D=4$ gauged supergravity coupled to abelian vector
multiplets (presence of U$(1)$ Fayet-Iliopoulos terms), give the general recipe
to construct supersymmetric solutions found in \cite{Cacciatori:2008ek}, and
simplify the equations of \cite{Cacciatori:2008ek} for the case where there is
essentially dependence on one coordinate only. This leads to the analogue of the
stabilization equations \cite{Ferrara:1996dd,Ferrara:1996um} in AdS space. In
section \ref{expl-ex} these equations are solved for various prepotentials,
amongst others for the STU model. This will lead to a variety of new BPS black
hole solutions with nonconstant scalars, whose physical properties are analyzed
as well. Generically, these black holes are solitonic in the sense that they have
no well-defined limit when the gauge coupling constants go to zero (at least not
in an obvious way). Finally, in section \ref{gen-nh} we determine the most
general supersymmetric static near-horizon geometry and obtain the attractor
equations in gauged supergravity. These are then solved in a simple example, and
the resulting moduli space for the scalars on the horizon is determined
explicitely.

\section{Supersymmetric black holes in ${\cal N}=2$, $D=4$ gauged supergravity}
\label{sugra}

We consider ${\cal N}=2$, $D=4$ gauged supergravity coupled to $n_V$ abelian
vector multiplets \cite{Andrianopoli:1996cm}\footnote{Throughout this paper,
we use the notations and conventions of \cite{Vambroes}.}.
Apart from the vierbein $e^a_{\mu}$, the bosonic field content includes the
vectors $A^I_{\mu}$ enumerated by $I=0,\ldots,n_V$, and the complex scalars
$z^{\alpha}$ where $\alpha=1,\ldots,n_V$. These scalars parametrize
a special K\"ahler manifold, i.~e.~, an $n_V$-dimensional
Hodge-K\"ahler manifold that is the base of a symplectic bundle, with the
covariantly holomorphic sections
\begin{equation}
{\cal V} = \left(\begin{array}{c} X^I \\ F_I\end{array}\right)\,, \qquad
{\cal D}_{\bar\alpha}{\cal V} = \partial_{\bar\alpha}{\cal V}-\frac 12
(\partial_{\bar\alpha}{\cal K}){\cal V}=0\,, \label{sympl-vec}
\end{equation}
where ${\cal K}$ is the K\"ahler potential and ${\cal D}$ denotes the
K\"ahler-covariant derivative. ${\cal V}$ obeys the symplectic constraint
\begin{equation}
\langle {\cal V}\,,\bar{\cal V}\rangle = X^I\bar F_I-F_I\bar X^I=i\,.
\end{equation}
To solve this condition, one defines
\begin{equation}
{\cal V}=e^{{\cal K}(z,\bar z)/2}v(z)\,,
\end{equation}
where $v(z)$ is a holomorphic symplectic vector,
\begin{equation}
v(z) = \left(\begin{array}{c} Z^I(z) \\ \frac{\partial}{\partial Z^I}F(Z)
\end{array}\right)\,.
\end{equation}
F is a homogeneous function of degree two, called the prepotential,
whose existence is assumed to obtain the last expression.
The K\"ahler potential is then
\begin{equation}
e^{-{\cal K}(z,\bar z)} = -i\langle v\,,\bar v\rangle\,.
\end{equation}
The matrix ${\cal N}_{IJ}$ determining the coupling between the scalars
$z^{\alpha}$ and the vectors $A^I_{\mu}$ is defined by the relations
\begin{equation}\label{defN}
F_I = {\cal N}_{IJ}X^J\,, \qquad {\cal D}_{\bar\alpha}\bar F_I = {\cal N}_{IJ}
{\cal D}_{\bar\alpha}\bar X^J\,.
\end{equation}
The bosonic action reads\footnote{We apologize for using the same letter
for the fluxes $F^I=dA^I$ and the lower part $F_I$ of the symplectic
section $\cal V$, but the meaning should be clear from the index position.}
\begin{eqnarray}
e^{-1}{\cal L}_{\text{bos}} &=& \frac 1{16\pi G}R + \frac 14(\text{Im}\,
{\cal N})_{IJ}F^I_{\mu\nu}F^{J\mu\nu} - \frac 18(\text{Re}\,{\cal N})_{IJ}\,e^{-1}
\epsilon^{\mu\nu\rho\sigma}F^I_{\mu\nu}F^J_{\rho\sigma} \nonumber \\
&& -g_{\alpha\bar\beta}\partial_{\mu}z^{\alpha}\partial^{\mu}\bar z^{\bar\beta}
- V\,, \label{action}
\end{eqnarray}
with the scalar potential
\eq
V = -2g^2\xi_I\xi_J[(\text{Im}\,{\cal N})^{-1|IJ}+8\bar X^IX^J]\,,
\feq
that results from U$(1)$ Fayet-Iliopoulos gauging. Here, $g$ denotes the
gauge coupling and the $\xi_I$ are constants. In what follows, we define
$g_I=g\xi_I$.

The most general timelike supersymmetric background of the theory described
above was constructed in \cite{Cacciatori:2008ek}, and is given by
\eq
ds^2 = -4|b|^2(dt+\sigma)^2 + |b|^{-2}(dz^2+e^{2\Phi}dwd\bar w)\ ,
\feq
where the complex function $b(z,w,\bar w)$, the real function $\Phi(z,w,\bar w)$
and the one-form $\sigma=\sigma_wdw+\sigma_{\bar w}d\bar w$, together with the
symplectic section \eqref{sympl-vec}\footnote{Note that also $\sigma$ and
$\cal V$ are independent of $t$.} are determined by the equations
\eq
\partial_z\Phi = 2ig_I\left(\frac{{\bar X}^I}b-\frac{X^I}{\bar b}\right)\ ,
\label{dzPhi}
\feq
\begin{eqnarray}
&&\qquad 4\partial\bar\partial\left(\frac{X^I}{\bar b}-\frac{\bar X^I}b\right) + \partial_z\left[e^{2\Phi}\partial_z
\left(\frac{X^I}{\bar b}-\frac{\bar X^I}b\right)\right]  \label{bianchi} \\
&&-2ig_J\partial_z\left\{e^{2\Phi}\left[|b|^{-2}(\text{Im}\,{\cal N})^{-1|IJ}
+ 2\left(\frac{X^I}{\bar b}+\frac{\bar X^I}b\right)\left(\frac{X^J}{\bar b}+\frac{\bar X^J}b\right)\right]\right\}= 0\,,
\nonumber
\end{eqnarray}
\begin{eqnarray}
&&\qquad 4\partial\bar\partial\left(\frac{F_I}{\bar b}-\frac{\bar F_I}b\right) + \partial_z\left[e^{2\Phi}\partial_z
\left(\frac{F_I}{\bar b}-\frac{\bar F_I}b\right)\right] \nonumber \\
&&-2ig_J\partial_z\left\{e^{2\Phi}\left[|b|^{-2}\text{Re}\,{\cal N}_{IL}(\text{Im}\,{\cal N})^{-1|JL}
+ 2\left(\frac{F_I}{\bar b}+\frac{\bar F_I}b\right)\left(\frac{X^J}{\bar b}+\frac{\bar X^J}b\right)\right]\right\}
\nonumber \\
&&-8ig_I e^{2\Phi}\left[\langle {\cal I}\,,\partial_z {\cal I}\rangle-
\frac{g_J}{|b|^2}\left(\frac{X^J}{\bar b}
+\frac{\bar X^J}b\right)\right] = 0\,, \label{maxwell}
\end{eqnarray}
\begin{equation}
2\partial\bar\partial\Phi=e^{2\Phi}\left[ig_I\partial_z\left(\frac{X^I}{\bar b}-\frac{\bar X^I}b\right)
+\frac2{|b|^2}g_Ig_J(\text{Im}\,{\cal N})^{-1|IJ}+4\left(\frac{g_I X^I}{\bar b}+\frac{g_I \bar X^I}b
\right)^2\right]\,, \label{Delta-Phi}
\end{equation}
\begin{equation}
d\sigma + 2\,\star^{(3)}\!\langle{\cal I}\,,d{\cal I}\rangle - \frac i{|b|^2}g_I\left(\frac{\bar X^I}b
+\frac{X^I}{\bar b}\right)e^{2\Phi}dw\wedge d\bar w=0\,. \label{dsigma}
\end{equation}
Here $\star^{(3)}$ is the Hodge star on the three-dimensional base with metric\footnote{Whereas
in the ungauged case, this base space is flat and thus has trivial holonomy, here we have U(1)
holonomy with torsion \cite{Cacciatori:2008ek}.}
\eq
ds_3^2 = dz^2+e^{2\Phi}dwd\bar w\ ,
\feq
and we defined $\partial=\partial_w$, $\bar\partial=\partial_{\bar w}$, as well as
\begin{equation}
{\cal I} = \text{Im}\left({\cal V}/\bar b\right)\ .
\end{equation}
Given $b$, $\Phi$, $\sigma$ and $\cal V$, the fluxes read
\begin{eqnarray}
F^I&=&2(dt+\sigma)\wedge d\left[bX^I+\bar b\bar X^I\right]+|b|^{-2}dz\wedge d\bar w
\left[\bar X^I(\bar\partial\bar b+iA_{\bar w}\bar b)+({\cal D}_{\alpha}X^I)b\bar\partial z^{\alpha}-
\right. \nonumber \\
&&\left. X^I(\bar\partial b-iA_{\bar w}b)-({\cal D}_{\bar\alpha}\bar X^I)\bar b\bar\partial\bar z^{\bar\alpha}
\right]-|b|^{-2}dz\wedge dw\left[\bar X^I(\partial\bar b+iA_w\bar b)+\right. \nonumber \\
&&\left.({\cal D}_{\alpha}X^I)b\partial z^{\alpha}-X^I(\partial b-iA_w b)-({\cal D}_{\bar\alpha}\bar X^I)
\bar b\partial\bar z^{\bar\alpha}\right]- \nonumber \\
&&\frac 12|b|^{-2}e^{2\Phi}dw\wedge d\bar w\left[\bar X^I(\partial_z\bar b+iA_z\bar b)+({\cal D}_{\alpha}
X^I)b\partial_z z^{\alpha}-X^I(\partial_z b-iA_z b)- \right.\nonumber \\
&&\left.({\cal D}_{\bar\alpha}\bar X^I)\bar b\partial_z\bar z^{\bar\alpha}-2ig_J
(\text{Im}\,{\cal N})^{-1|IJ}\right]\,. \label{fluxes}
\end{eqnarray}
In \eqref{fluxes}, $A_{\mu}$ is the gauge field of the K\"ahler U$(1)$,
\eq
A_{\mu} = -\frac i2(\partial_{\alpha}{\cal K}\partial_{\mu}z^{\alpha} -
         \partial_{\bar\alpha}{\cal K}\partial_{\mu}{\bar z}^{\bar\alpha})\,.
\feq

In order to solve the system \eqref{dzPhi}-\eqref{dsigma} we shall assume
that $b$ and $\cal V$ depend on the coordinate $z$ only, and use the
separation ansatz $\Phi=\psi(z)+\gamma(w,\bar w)$. Furthermore, we are
looking for static solutions, i.e., $\sigma=0$. Then \eqref{dsigma}
boils down to
\eq
\langle{\cal I}\,,\partial_z{\cal I}\rangle = |b|^{-2}g_I\left(\frac{\bar X^I}b
+ \frac{X^I}{\bar b}\right)\,. \label{dsigma'}
\feq
Using this, one can integrate \eqref{maxwell} once, with the result
\begin{eqnarray}
e^{2\psi}\partial_z\left(\frac{F_I}{\bar b} - \frac{\bar F_I}b\right)
&-&2ig_Je^{2\psi}\left[|b|^{-2}\text{Re}\,{\cal N}_{IL}(\text{Im}\,
{\cal N})^{-1|JL}\right. \nonumber \\
&+& 2\left.\left(\frac{F_I}{\bar b}+\frac{\bar F_I}b\right)\left(\frac{X^J}
{\bar b}+\frac{\bar X^J}b\right)\right] = -4\pi iq_I\,, \label{stab1}
\end{eqnarray}
while \eqref{bianchi} yields
\begin{eqnarray}
e^{2\psi}\partial_z\left(\frac{X^I}{\bar b} - \frac{\bar X^I}b\right)
&-&2ig_Je^{2\psi}\left[|b|^{-2}(\text{Im}\,{\cal N})^{-1|IJ}\right. \nonumber \\
&+& 2\left.\left(\frac{X^I}{\bar b}+\frac{\bar X^I}b\right)\left(\frac{X^J}
{\bar b}+\frac{\bar X^J}b\right)\right] = -4\pi ip^I\,. \label{stab2}
\end{eqnarray}
Here, $q_I$ and $p^I$ denote integration constants that will be identified
below with the electric and magnetic charge densities respectively.
Finally, \eqref{dzPhi} and \eqref{Delta-Phi} reduce to
\eq
\partial_z\psi = 2ig_I\left(\frac{{\bar X}^I}b-\frac{X^I}{\bar b}\right)\,,
\label{dzpsi}
\feq
\eq
-4\partial\bar\partial\gamma = \kappa e^{2\gamma}\,, \qquad
\kappa = -8\pi g_Ip^I\,, \label{liouville}
\feq
where we used the contraction of \eqref{stab2} with $g_I$. \eqref{liouville}
is the Liouville equation and implies that the metric $e^{2\gamma}dwd\bar w$
has constant curvature $\kappa$, determined by the magnetic charges $p^I$.
In the following section, we shall solve the system
\eqref{dsigma'}-\eqref{liouville} explicitely for various prepotentials.

\section{Explicit examples}
\label{expl-ex}

\subsection{Prepotential $F=-iX^0X^1$}
\label{X0X1}

Let us now solve the above equations for the SU(1,1)/U(1) model with prepotential
$F=-iX^0X^1$, that has $n_V=1$ (one vector multiplet), and thus just one complex scalar $\tau$.
Choosing $Z^0=1$, $Z^1=\tau$, the symplectic vector $v$ becomes
\eq
v = \left(\begin{array}{c} 1 \\ \tau \\ -i\tau \\ -i\end{array}\right)\ .
\label{v-X0X1}
\feq
The K\"ahler potential, metric and kinetic matrix for the vectors are given
respectively by
\eq
e^{-{\cal K}} = 2(\tau + \bar\tau)\ , \qquad g_{\tau\bar\tau} = \partial_\tau\partial_{\bar\tau}
{\cal K} = (\tau + \bar\tau)^{-2}\ ,
\feq
\eq
{\cal N} = \left(\begin{array}{cc} -i\tau & 0 \\ 0 & -\frac i\tau\end{array}\right)\ .
\feq
Note that positivity of the kinetic terms in the action requires
${\mathrm{Re}}\tau>0$. For the scalar potential one obtains
\eq
V = -\frac4{\tau+\bar\tau}(g_0^2 + 2g_0g_1\tau + 2g_0g_1\bar\tau
+ g_1^2\tau\bar\tau)\ , \label{pot_su11}
\feq
which has an extremum at $\tau=\bar\tau=|g_0/g_1|$. In what follows we assume
$g_I>0$. The K\"ahler U(1) is
\eq
A_{\mu} = \frac i{2(\tau+\bar\tau)}\partial_{\mu}(\tau-\bar\tau)\ .
\feq
In order to solve the system \eqref{dsigma'}-\eqref{liouville} we shall take
$\tau=\bar\tau$ (this includes the extremum of the potential, and thus the
AdS vacuum). This implies $\mbox{Re}\,{\cal N}=0$. Furthermore, we assume that
$b$ is imaginary,
\eq
b=iN(z)\ , \qquad N\ \mbox{real}\ .
\feq
Then one has
\eq
\frac{X^I}{\bar b} + \frac{{\bar X}^I}b = \frac{F_I}{\bar b} -
\frac{{\bar F}_I}b = 0\ , \qquad \langle{\cal I}\,,d{\cal I}\rangle = 0\ ,
\feq
hence \eqref{dsigma'} is trivially satisfied and \eqref{stab1} is solved
for $q_I=0$. Defining
\eq
H^0 = \frac{2X^0}N = \frac1{N\sqrt\tau}\ , \qquad H^1 = \frac{2X^1}N =
\frac{\sqrt{\tau}}N\ ,
\feq
equ.~\eqref{stab2} leads to
\eq
e^{2\psi}\left[\frac12\partial_zH^I + g_I(H^I)^2\right] = -2\pi p^I\ , \qquad
\mbox{no summation over $I$}\ . \label{stab2'}
\feq
Inspired by the minimal case \cite{Cacciatori:2004rt}, we make the ansatz
\eq
\psi = \ln(az^2+c)\ , \qquad H^I = \frac{\alpha^Iz+\beta^I}{az^2+c}\ ,
\label{ans-psi-H}
\feq
with $a,c,\alpha^I,\beta^I\in\bR$ constants. Then the
remaining equations \eqref{dzpsi} and \eqref{stab2'} are satisfied iff
\eq
\alpha^I = \frac a{2g_I}\ , \qquad
\frac{ac}{4g_I} + g_I(\beta^I)^2 + 2\pi p^I= 0\ , \qquad g_I\beta^I = 0\ .
\label{rel-coeff}
\feq
Note that these relations imply also
\eq
g_0p^0 = g_1p^1\ . \label{p0p1}
\feq
The scalar field and lapse function are respectively given by
\eq
\tau = \frac{H^1}{H^0} = \frac{g_0}{g_1}\,\frac{az-2g_0\beta^0}{az+2g_0\beta^0}\ ,
\qquad N^2 = (H^0H^1)^{-1} = \frac{4g_0g_1(az^2+c)^2}{a^2z^2-4(g_0\beta^0)^2}\ .
\feq
In what follows, we shall assume $a>0$, $c<0$. Then the solution will have
an event horizon at $z=z_h=\sqrt{-c/a}$. The scalar $\tau$ is positive as
long as $z>2|g_0\beta^0|/a$. We want the dangerous point $z=2|g_0\beta^0|/a$
to be hidden behind the horizon, i.e., $z_h>2|g_0\beta^0|/a$, which, by
using the second relation of \eqref{rel-coeff}, implies $p^0>0$. By \eqref{p0p1}
we have then also $p^1>0$ and thus $\kappa<0$, so that the horizon geometry must
be hyperbolic. Notice that for $\beta^0=0$, the scalar field is constant, and
the solution reduces to the one of minimal gauged supergravity discovered in
\cite{Caldarelli:1998hg}.

The above black hole geometry has two scaling symmetries, namely
\eq
(t,z,w,a,c,\beta^I,\kappa) \mapsto (t/\lambda,\lambda z,\lambda w,a/\lambda^2,
c,\beta^I/\lambda,\kappa/\lambda^2)\ , \label{scale1}
\feq
and
\eq
(t,z,w,a,c,\beta^I,\kappa) \mapsto (t/\lambda,\lambda z,w,a/\lambda,
\lambda c,\beta^I,\kappa)\ . \label{scale2}
\feq
One can use the first to set $\kappa=-1$ and then the second (that leaves
$\kappa$ invariant) to choose $a=1$. If we define the parameter $\nu$ by
$\sinh\nu=2\sqrt2g_0\beta^0$, the line element reads
\eq
ds^2 = -4N^2dt^2 + \frac{dz^2}{N^2} + \frac{z^2-\frac12\sinh^2\nu}{4g_0g_1}
e^{2\gamma}dwd\bar w\ , \label{metr-2charge}
\feq
where
\eq
N^2 = \frac{4g_0g_1(z^2-\frac12\cosh^2\nu)^2}{z^2-\frac12\sinh^2\nu}\ .
\feq
The scalar and the fluxes become
\eq
\tau = \frac{g_0}{g_1}\,\frac{z-\frac1{\sqrt2}\sinh\nu}{z+\frac1{\sqrt2}
\sinh\nu}\ , \qquad F^I = 2\pi ip^Ie^{2\gamma}dw\wedge d\bar w\ .
\label{flux-2charge}
\feq
This yields for the magnetic charges $P^I$
\eq
P^I = \frac1{4\pi}\int F^I = p^IV\ , \qquad V \equiv \frac i2\int e^{2\gamma}
dw\wedge d\bar w\ , \label{vol}
\feq
confirming that the $p^I$ represent the magnetic charge densities. Note that
these obey a Dirac quantization condition: From $\kappa=-8\pi g_Ip^I=-1$ and
\eqref{p0p1} we get
\eq
p^I = \frac1{16\pi g_I}\ ,
\feq
i.e., the $p^I$ are quantized in terms of the inverse coupling constants.
The entropy density of the one-parameter solution
\eqref{metr-2charge}-\eqref{flux-2charge} is
\eq
s = \frac SV = 8\pi^2p^0p^1\ . \label{entr-X0X1}
\feq

\subsection{STU model with $F=-2\sqrt{-X^0X^1X^2X^3}$}

Next we consider the STU model with prepotential
\eq
F = -2\sqrt{-X^0X^1X^2X^3}\ .
\feq
Choosing $Z^0=1$, $Z^1=\tau^2\tau^3$, $Z^2=\tau^1\tau^3$, $Z^3=\tau^1\tau^2$,
the symplectic vector $v$ becomes
\eq
v = (1,\tau^2\tau^3,\tau^1\tau^3,\tau^1\tau^2,-i\tau^1\tau^2\tau^3,-i\tau^1,
-i\tau^2,-i\tau^3)^T\ .
\feq
The K\"ahler potential and metric are given respectively by
\eq
e^{-{\cal K}} = 8\,\mbox{Re}\tau^1\mbox{Re}\tau^2\mbox{Re}\tau^3\ ,
\label{Kaehler-stu}
\feq
\eq
g_{\alpha\bar\alpha} = g_{\bar\alpha\alpha} = (\tau^{\alpha} +
\bar\tau^{\bar\alpha})^{-2}\ , \qquad \alpha =1,2,3\ , \label{metr-stu}
\feq
and all other components vanishing. In what follows, we assume $\tau^{\alpha}$
real and positive. Then the kinetic matrix for the vectors is
\eq
{\cal N} = -i\,\mbox{diag}(\tau^1\tau^2\tau^3,\frac{\tau^1}{\tau^2\tau^3},
\frac{\tau^2}{\tau^1\tau^3},\frac{\tau^3}{\tau^1\tau^2})\ ,
\feq
and hence $\mbox{Re}\,{\cal N}=0$. Notice also that
\eq
(\mbox{Im}\,{\cal N})^{-1} = -8\,\mbox{diag}((X^0)^2,(X^1)^2,(X^2)^2,(X^3)^2)\ .
\feq
For the scalar potential one obtains
\eq
V = -4\left(\frac{g_0g_1}{\tau^1}+g_2g_3\tau^1+\frac{g_0g_2}{\tau^2}+g_1g_3\tau^2
+\frac{g_0g_3}{\tau^3}+g_1g_2\tau^3\right)\ , \label{pot_stu}
\feq
which has an extremum at
\[
\tau^1 = \left(\frac{g_0g_1}{g_2g_3}\right)^{1/2}\ , \qquad
\tau^2 = \left(\frac{g_0g_2}{g_1g_3}\right)^{1/2}\ , \qquad
\tau^3 = \left(\frac{g_0g_3}{g_1g_2}\right)^{1/2}\ .
\]
Note that for all $g_I$ equal, this model can be embedded into ${\cal N}=8$ gauged
supergravity as well \cite{Duff:1999gh}.

In order to solve the equations \eqref{dsigma'}-\eqref{dzpsi}, we use
once more the assumption $b=iN(z)$ for $N$ real. Since the $F_I$ are imaginary
and $X^I$ real, \eqref{dsigma'} is identically satisfied, \eqref{stab1} gives
$q_I=0$ and \eqref{stab2} simplifies to
\eq
e^{2\psi}\left[\partial_zH^I + 8g_I(H^I)^2\right] = -2\pi p^I\ , \qquad
\mbox{no summation over $I$}\ , \label{stab2-stu}
\feq
where $H^I\equiv X^I/N$. As before, we employ the ansatz \eqref{ans-psi-H}
for the functions $\psi$ and $H^I$. Then \eqref{dzpsi} and \eqref{stab2-stu}
are fulfilled if the following constraints hold:
\eq
\alpha^I = \frac a{8g_I}\ , \qquad
\frac{ac}{8g_I} + 8g_I(\beta^I)^2 + 2\pi p^I= 0\ , \qquad g_I\beta^I = 0\ .
\label{rel-coeff-stu}
\feq
The scalars fields and the lapse function read
\eq
\tau^{\alpha} = \frac1{8X^0X^{\alpha}} = \frac{(H^0H^1H^2H^3)^{1/2}}{H^0H^{\alpha}}\ ,
\feq
\eq
N^2 = \frac18(H^0H^1H^2H^3)^{-1/2} = \frac{(az^2+c)^2}{8\prod_I(\alpha^Iz+
\beta^I)^{1/2}}\ . \label{lapse-stu}
\feq
For the line element and the fluxes one gets respectively
\eq
ds^2 = -4N^2dt^2 + \frac{dz^2}{N^2} + 8\prod_{I=0}^3(\alpha^Iz+
\beta^I)^{\frac 12} e^{2\gamma}dwd\bar w\,,
\feq
\eq
F^I= 2\pi i p^I e^{2\gamma} dw\wedge d\bar w\,,
\feq
so that the $p^I$ represent again the magnetic charge densities.
In what follows, we shall assume $g_I>0$, $a>0$ (and thus $\alpha^I>0$ by
\eqref{rel-coeff-stu}), as well as $c<0$, so that there is a horizon at
$z=z_h=\sqrt{-c/a}$. The entropy density can be written in the form
\eq
\frac SV = 2\prod_{I=0}^3\left(\beta^I+\sqrt{\frac{\pi p^I}{4g_I}+(\beta^I)^2}
\right)^{\frac 12}\,. \label{entr-stu1}
\feq
The solution is again invariant under the scaling symmetries \eqref{scale1},
\eqref{scale2} that allow to set $a=1$, $\kappa=0,\pm 1$ without loss of
generality.

We must ensure that the moduli $\tau^{\alpha}$ be positive in the whole region
outside the horizon. This is guaranteed if $z_h>-\beta^I/\alpha^I$ $\forall I$.
A sufficient condition for this is $-c>(\beta^I/\alpha^I)^2$, which,
by using the second relation of \eqref{rel-coeff-stu}, yields $p^I>0$ and thus
$\kappa<0$, so that the horizon geometry is hyperbolic in this case. But the
condition $-c>(\beta^I/\alpha^I)^2$ is not necessary in general and for
suitable choices of the parameters, other geometries are allowed. The constraint
$g_I\beta^I=0$ with $g_J>0$ shows that at least one of the $\beta^I$
must be negative. It is easy to show that if only one of them is negative, then
necessarily $\kappa<0$. Indeed, let us assume to be $\beta^0$ the only negative
coefficient. Then, we need to impose only one condition, which is equivalent to
\eq
-c>(\beta^0)^2/(\alpha^0)^2 = 64(g_0\beta^0)^2\,.
\feq
Then
\eqn
&& -\frac \kappa4=2\pi g_I p^I= -\frac c2 -8\sum_I g_I^2 (\beta^I)^2>
-\frac c2 -8g_0^2 (\beta^0)^2- 8(\sum_{i=1}^3 g_i \beta^i)^2 \cr
&& \qquad =-\frac c2 -16 g_0^2 (\beta^0)^2>16 g_0^2 (\beta^0)^2>0\,.
\feqn
Let us the consider the opposite situation, when $\beta^0>0$ is positive and
assume for simplicity $g_i=g$, $\beta^i=-\beta$, for $i=1,2,3$ and $g$ and
$\beta$ positive. Then $g_0\beta^0=3g\beta$. The singularities are then hidden
by the horizon if and only if $c+64g^2\beta^2<0$. Now
\eq
\frac\kappa4=-2\pi g_I p^I =\frac c2 +8\sum_I{g_I^2(\beta^I)^2} =
\frac12(c+192 g^2\beta^2)\,.
\feq
Then, a spherical topology ($\kappa=1$) for the horizon is admitted if, in
this example, the parameters satisfy
\eq
-c = 192g^2\beta^2-\frac12\,.
\feq
Combining this with $-c>64g^2\beta^2$ yields $16g\beta>1$. A flat horizon
($\kappa=0$) appears for $-c=192 g^2\beta^2$.

In order to say more about the allowed horizon geometries, we shall now solve
the equations \eqref{rel-coeff-stu} systematically. To this end,
note that once we fix $g_I$ and the charge densities $p^I$ (taking into
account the condition $8\pi g_I p^I=-\kappa$), all other parameters are
generically determined by the relations (\ref{rel-coeff-stu}). Indeed, defining
$y_I=g_I\beta^I$ (no summation over $I$), and using $\alpha^I=1/(8g_I)$
together with
\eq
c=\frac \kappa2-16 \sum_I g_I^2(\beta^I)^2\,, \label{cfactor}
\feq
one is left with four equations for the four unknowns $y_I$:
\eqn
&& 3y_1^2-y_2^2-y_3^2-y_0^2=-\phi_1\,, \label{phi1}\\
&& -y_1^2+3y_2^2-y_3^2-y_0^2=-\phi_2\,,\label{phi2}\\
&& -y_1^2-y_2^2+3y_3^2-y_0^2=-\phi_3\,,\label{phi3}\\
&& y_0+y_1+y_2+y_3=0\,,\label{plane}
\feqn
where $\phi_I=\frac\kappa{32}+\pi g_Ip^I$ (no summation over $I$) satisfying
$\sum_I\phi_I=0$. The solutions are the intersections between
three hypersurfaces of degree 2 in $\mathbb{R}^4$ and a hyperplane, so that
we expect generically a maximum of eight isolated points. However, it can
happen that these four hypersurfaces do not intersect transversally so that
there is a higher-dimensional intersection. This happens when the determinant
of gradients vanishes,
\eq
0={\rm det}\left(
\begin{array}{cccc}
3y_1 & -y_2 & -y_3 & -y_0 \\
-y_1 & 3y_2 & -y_3 & -y_0 \\
-y_1 & -y_2 & 3y_3 & -y_0 \\
1 & 1 & 1 & 1
\end{array}
\right)=16(y_1 y_2 y_3 +y_0 y_1 y_2 +y_0 y_1 y_3 +y_0 y_2 y_3)\,.
\feq
Using the hyperplane equation, the degeneracy condition takes the form
\eq
0=(y_0 y_3-y_1 y_2)(y_0+y_3)=(y_0+y_3) {\rm det}\left(\begin{array}{cc}
y_0 & y_1\\ y_2 & y_3 \end{array} \right)\,.
\feq
Then, either $y_0+y_3=y_1+y_2=0$, or there exists some $\lambda\in\mathbb{R}$
such that $(y_0, y_1)=\lambda (y_2, y_3)$. Inserted into the hyperplane
equation this gives
\eq
(1+\lambda) (y_2+y_3)=0\,,
\feq
which yields $y_2+y_3=y_0+y_1=0$ or $\lambda=-1$. In conclusion, we see that
the degeneracy conditions are equivalent to
\eq
y_{\sigma(1)}+y_{\sigma(2)}=y_{\sigma(3)}+y_{\sigma(0)}=0\,, \label{degenercy}
\feq
where $\sigma$ is a fixed element in the symmetric group $S_4$\footnote{There
is a possible degeneracy for each choice of $\sigma$, however only three of
them are indeed distinct.}. We see that (\ref{degenercy}) is compatible with
the hyperplane equation. However, the degeneracy appears only when it is
compatible with the whole system. Substituting into the system we see that this
happens when
\eq
\phi_{\sigma(1)}=\phi_{\sigma(2)}=-\phi_{\sigma(3)}=-\phi_{\sigma(0)}\,,
\feq
which means
\eqn
&& g_{\sigma(1)}p^{\sigma(1)}=g_{\sigma(2)}p^{\sigma(2)}\,, \qquad\ g_{\sigma(0)}
p^{\sigma(0)}=g_{\sigma(3)}p^{\sigma(3)}\,,\cr
&& g_{\sigma(1)}p^{\sigma(1)}=-g_{\sigma(0)}p^{\sigma(0)} +\frac 1{16\pi}\,.
\feqn
When these conditions are satisfied, we see that a free parameter is left by
(\ref{rel-coeff-stu}), let us say $\beta^0$. Note that the conditions imply
$\kappa=-1$. The entropy density \eqref{entr-stu1} boils down to
\eq
\frac SV = \frac{\pi}2\sqrt[4] {\prod_{I=0}^3 \frac {p^I}{g_I}}\,,
\feq
which is thus independent of the free parameter.

Let us now study the isolated points. We will see that these allow for
solutions with $\kappa\ge 0$. To this aim let us set $Z_I=y_I^2$, $I=0,1,2,3$.
Then the equations (\ref{phi1}), (\ref{phi2}) and (\ref{phi3})
can be solved for $Z_i$, $i=1,2,3$ as functions of $Z_0$:
\eqn
Z_i =Z_0 +\frac 14 (\phi_0-\phi_i)\,. \label{function}
\feqn
Equation (\ref{plane}) becomes
\eqn
\sqrt {Z_0}+\sigma_1 \sqrt{Z_1}=\sigma_2 \sqrt{Z_2} +\sigma_3 \sqrt{Z_3}\,,
\feqn
where $\sigma_i$ are signs. Taking the square of this relation and using
(\ref{function}) we get\footnote{We often use tacitly the relation
$\phi_0+\ldots +\phi_3=0$.}
\eqn
2\sigma_2\sigma_3\sqrt{Z_2 Z_3}-2\sigma_1\sqrt{Z_0 Z_1}=Z_0+Z_1-Z_2-Z_3=
-\frac 12 (\phi_2+\phi_3)\,,
\feqn
that is
\eqn
\sigma_1\sigma_2\sigma_3\sqrt{Z_2 Z_3}=\sqrt{Z_0 Z_1}-\frac14\sigma_1
(\phi_2+\phi_3)\,.
\feqn
Squaring this and making use of
\eqn
Z_2 Z_3 -Z_0 Z_1= \frac1{16}(\phi_0-\phi_2)(\phi_0-\phi_3)-\frac12 Z_0
(\phi_2+\phi_3)\,,
\feqn
one obtains
\eqn
\sqrt{Z_0 Z_1}=-Z_0 + \frac{(\phi_0-\phi_2)(\phi_0-\phi_3)}{(\phi_2+\phi_3)}
-\frac18 (\phi_2+\phi_3)\,.
\feqn
Here we assumed $\phi_2+\phi_3\neq 0$, otherwise we fall in the degenerate
case. Taking the square and using the expression for $Z_1$ as a function of
$Z_0$ we finally get
\eqn
Z_I= -\frac{(4\phi_I^2-\sum_{J=0}^3\phi_J^2)^2}
{64\prod_{J\neq I}(\phi_J+\phi_I)}\,.
\feqn 
The acceptable solutions are the ones satisfying the condition
\eqn
\prod_{J\neq I} (\phi_J+\phi_I)<0\,. \label{positiv}
\feqn
Note that this condition is the same for all $I$: indeed
$\prod_{J\neq I}(\phi_J+\phi_I)$ does not depend on $I$ because of the identity
$\sum_J \phi_J=0$. Then
\eqn
y_I = \frac{\sigma_I(4\phi_I^2-\sum_{J=0}^3\phi_J^2)}{8\sqrt{-\prod_{J\neq I}
(\phi_J+\phi_I)}}\,, \label{soluz}
\feqn
for certain signs $\sigma_I$. Notice that the denominator is the same for all
$Z_I$. By inspection we see that all signs must be equal as well. Indeed, if
this is the case, we immediately see that (\ref{plane}) is satisfied. Moreover,
if $\{Y_I\}$ is a solution then $\{-Y_I\}$ is another solution and we can
consider only the cases when there is just one negative sign or two negative
signs. In the first case, without loss of generality, we can assume that only
$y_3$ is negative, so that $\sigma_3=-1$ and $\sigma_i=1$ for $i\neq 3$.
Plugging (\ref{soluz}) into (\ref{plane}) we get
$3\phi_3^2-\phi_0^2-\phi_1^2-\phi_2^2=0$ which is equivalent to $y_3=0$
and then there is not a real different choice of the sign.\\
Doing the same for the case of two negative signs, for example
$\sigma_2=\sigma_3=-1$, we find $\phi_0^2+\phi_1^2-\phi_2^2-\phi_3^2=0$ that
is equivalent to $0=\phi_2\phi_3-\phi_1\phi_0$ which is the degeneracy
condition. Then we conclude that
\eqn
y_I = \pm\frac{4\phi_I^2-\sum_{J=0}^3\phi_J^2}{8\sqrt{-\prod_{J\neq I}
(\phi_J+\phi_I)}}\,, \label{soluzfin}
\feqn
where the sign is the same for all $I$.\\
Let us suppose to have chosen the charges and the coupling constants such that
$\sum_I g_I p^I=-\kappa/8\pi$. Then, the consistency condition (\ref{positiv})
takes the form
\begin{displaymath}
\left(\frac{\kappa}{16\pi} + (g_2 p^2)^2 +(g_3 p^3)^2\right)\left(\frac{\kappa}
{16\pi} + (g_1 p^1)^2 +(g_3 p^3)^2\right)\left(\frac{\kappa}{16\pi} +
(g_1 p^1)^2 + (g_2 p^2)^2\right) > 0\,.
\end{displaymath}
Note that a general choice of the possible parameters does not allow for a
horizon. The existence condition requires $y_I>0$ when $p_I<0$.\\
As an example, let us consider the case $p_i>0$. Then, we can look for
solutions with $\kappa\geq0$. As $c$ must be negative to have a horizon, we
see from (\ref{cfactor}) that
\eqn
0\leq \kappa < 32 \sum_{I=0}^3 y_I^2\,.
\feqn 
The solutions are given by (\ref{soluzfin}). As $p^0=-p^1-p^2-p^3-\kappa/8\pi$
is the only negative charge, we have to be careful with the
sign of $y_0$ only. But $\phi_i>0$ so that
$3(\sum_{i=1}^3\phi_i)^2-\sum_{i=1}^3 \phi_i^2>0$ and then $y_0$ is positive if
we chose the plus sign in (\ref{soluzfin}). This provides the desired solution.

Notice that, if all $g_I$ are equal, the 4-charge black holes found in this
subsection can be uplifted to 11 dimensions along the lines
of \cite{Cvetic:1999xp}. That might be interesting to do.

\subsection{STU model with $F=-X^1X^2X^3/X^0$}

Another interesting model is the one with
prepotential $F=-X^1X^2X^3/X^0$. In the ungauged case, this is related by a
symplectic transformation to the model with $F=-2(-X^0X^1X^2X^3)^{1/2}$
considered above \cite{Bellucci:2008cb}.
However, in the presence of gauging, symplectic covariance is broken, so that
this prepotential will lead to different physics.

Choosing $Z^0=1$, $Z^{\alpha}=i\tau^{\alpha}$, $\alpha=1,2,3$,
the symplectic vector $v$ becomes
\eq
v = (1,i\tau^1,i\tau^2,i\tau^3,-i\tau^1\tau^2\tau^3,\tau^2\tau^3,
\tau^1\tau^3,\tau^1\tau^2)^T\ .
\feq
The K\"ahler potential and metric are again given by \eqref{Kaehler-stu}
and \eqref{metr-stu} respectively. In the following, we assume $\tau^{\alpha}$
real and positive ("vanishing axions" condition). Then the kinetic matrix for
the vectors is
\eq
{\cal N} = -i\,\mbox{diag}(\tau^1\tau^2\tau^3,\frac{\tau^2\tau^3}{\tau^1},
\frac{\tau^1\tau^3}{\tau^2},\frac{\tau^1\tau^2}{\tau^3})\ ,
\feq
and thus $\mbox{Re}\,{\cal N}=0$. Notice also that
\eq
(\mbox{Im}\,{\cal N})^{-1} = 8\,\mbox{diag}(-(X^0)^2,(X^1)^2,(X^2)^2,(X^3)^2)\ .
\feq
For the scalar potential one gets
\eq
V = -4\left(\frac{g_2g_3}{\tau^1}+\frac{g_1g_3}{\tau^2}+\frac{g_1g_2}{\tau^3}
\right)\ , \label{pot_nocrit}
\feq
which has no critical point, so that there are no AdS$_4$ vacua with constant
moduli.

In what follows we choose $b$ real. Since $X^0$ and $F_{\alpha}$
($\alpha=1,2,3$) are real as well, and $F_0$, $X^{\alpha}$ are
imaginary, one has
\eq
\frac{X^{\alpha}}{\bar b} + \frac{\bar X^{\alpha}}b = \frac{X^0}{\bar b} -
\frac{\bar X^0}b = 0\,, \qquad \frac{F_{\alpha}}{\bar b} -
\frac{\bar F_{\alpha}}b = \frac{F_0}{\bar b} + \frac{\bar F_0}b = 0\,,
\feq
and hence $\langle{\cal I}\,,d{\cal I}\rangle=0$. If we make the choice
$g_0=0$\footnote{Note that this does not affect the scalar
potential \eqref{pot_nocrit}.}, equ.~\eqref{dsigma'} holds identically.
\eqref{stab1} for $I=\alpha$ and \eqref{stab2} for $I=0$ are also
satisfied for $q_{\alpha}=p^0=0$. On the other hand, defining
$H^{\alpha}=X^{\alpha}/b$, $H^0=1/(X^0b)$, the remaining equations of
\eqref{stab1} and \eqref{stab2} boil down to
\eq
e^{2\psi}\partial_zH^0 = 16\pi q_0\,, \label{dzH0}
\feq
\eq
e^{2\psi}\left[\partial_zH^{\alpha} - 8ig_{\alpha}(H^{\alpha})^2\right] =
-2\pi ip^{\alpha}\,, \qquad
\text{no summation over $\alpha$}\,, \label{Halpha}
\feq
whereas \eqref{dzpsi} gives
\eq
\partial_z\psi = -4ig_{\alpha}H^{\alpha}\,. \label{psi'}
\feq
Plugging \eqref{psi'} into \eqref{Halpha} yields
\eq\label{psi''}
e^{2\psi}\left[\psi'' + \frac 23{\psi'}^2\right] = -24\pi gp\,,
\feq
where the prime indicates a derivative w.r.t.~$z$, and we made the further
assumption $g_\alpha p^\alpha=gp$ and $g_\alpha H^\alpha=gH$ for $\alpha=1,2,3$
(no summation over $\alpha$). This is equivalent to taking $g_\alpha \tau^\alpha =g\tau$,
$\alpha=1,2,3$. Setting $y=e^{2\psi/3}$, (\ref{psi''}) can be rewritten as
\eq
y''=-\frac{16\pi gp}{y^2}\,,
\feq
and thus
\eq\label{monotonic}
y'=\pm \sqrt {C+\frac {32\pi gp}y}\,,
\feq
with $C$ an integration constant. We can chose the upper sign, the other one corresponding to
the inversion $z\leftrightarrow -z$. From this equation we see that the relation between $z$ and
$y$ is monotonic and we can use $y$ in place of $z$ as a new coordinate.
Then (\ref{Halpha}) takes the form
\eq
\sqrt{C+\frac {32\pi gp}y} \frac{dH}{dy}-8ig H^2=-2\pi i \frac p{y^3}\,.
\feq
This is a Riccati equation with particular solution
\eq
\tilde H(y)=\frac i{8gy} \sqrt{C+\frac {32\pi gp}y}\,.
\feq
The general solution is then
\eq
H(y)=\tilde H(y)+\frac{iK}{y^2\left[1-\frac K{2\pi p} \sqrt{C+\frac {32\pi gp}y}\right]}\,,
\feq
where $K$ denotes another integration constant. However, using $\psi=\frac 32 \ln y$ and
(\ref{psi'}) we see that $K=0$ so that
\eq
H^\alpha(y)=\frac i{8g_\alpha y} \sqrt{C+\frac {32\pi gp}y}\,.
\feq 
In the same way we can solve (\ref{dzH0}), with the result
\eq
H^0(y)=\frac {q_0}{32\pi g^2 p^2} \sqrt{C+\frac {32\pi gp}y} \left( \frac 23 C -\frac {32\pi gp}{3y} \right) +h^0 \,.
\feq
The function $b$ is given by
\eq
b^4 = -i(8H^0H^1H^2H^3)^{-1} =\frac {64 g_1 g_2 g_3 y^3}{H^0 (y) \left( C+\frac {32\pi gp}y \right)^{3/2}} \,.
\label{bb^4}
\feq
The scalar fields $\tau^\alpha$ are obtained from $X^0=1/(H^0b)=e^{{\cal K}/2}$,
yielding
\eq
\tau^\alpha = \frac {\sqrt {g_1 g_2 g_3}}{g_\alpha} \frac {(yH^0)^\frac 12}{\left(C+\frac {32\pi gp}y\right)^\frac 14}\,.
\feq
For the metric and the fluxes one gets
\eq
ds^2 = -4b^2dt^2 + b^{-2}\left(\frac {dy^2}{C+\frac {32\pi gp}y} + y^3 e^{2\gamma}dwd\bar w\right)\,,
\label{metr-stu2}
\feq
\eq
F^0 = 4dt\wedge d(H^0)^{-1}\,, \qquad F^{\alpha} = 2\pi ip^{\alpha}e^{2\gamma}
dw\wedge d\bar w\,,
\feq
where $\gamma$ satisfies
\eq
\partial\bar\partial\gamma = 6\pi gpe^{2\gamma}\,.
\feq
The solution carries thus one electric and three magnetic charges, namely
\eq
Q_0 = q_0V\,, \qquad P^{\alpha} = p^{\alpha}V\,,
\feq
with $V$ given in \eqref{vol}.
Note that $e^{2\psi}>0$ corresponds to $y>0$. Then, there is an event horizon at $y=0$,
if $p>0$\footnote{We have chosen also $g>0$.}, and thus (cf.~\eqref{liouville})
$\kappa=-24\pi gp<0$, so that the horizon is hyperbolic. If we assume also $C\geq 0$,
$q_0<0$, and $h^0> |q_0|C^{3/2}/(48\pi g^2 p^2)$, the scalar fields are real and positive for
$0\le y<\infty$. The values of the scalars on the horizon and the entropy are
\eq
\tau^\alpha_{\text{hor}}=\frac {\sqrt {g_1 g_2 g_3}}{g_\alpha}\left(\frac {|q_0|}{3gp}\right)^{1/2} =
\frac{(g_1g_2g_3)^{1/3}}{g_{\alpha}}\frac{|Q_0/3|^{1/2}}{(P^1P^2P^3)^{1/6}}\,,
\feq
\eq
S=\pi V\left(\frac{|q_0|gp}{3g_1g_2g_3}\right)^{1/2} = \frac{\pi |Q_0/3|^{1/2}(P^1P^2P^3)^{1/6}}
{(g_1g_2g_3)^{1/3}}\,.
\feq
We see that in this case both $\tau^\alpha_{\text{hor}}$ and $S$ depend on the charges only.
The solution \eqref{metr-stu2} interpolates between AdS$_2$$\times$H$^2$ near the
horizon and a curved domain wall for $y\to\infty$. The worldvolume of the domain wall is
the open Einstein static universe $\bR$$\times$H$^2$.

\subsection{Prepotential $F=\frac i4 X^I \eta_{IJ} X^J$}
Let us now consider the SU$(1,n)$/(U(1)$\times$ SU$(n)$) model with prepotential
$F=\frac i4 X^I\eta_{IJ}X^J$, where the scalar fields are $X^I$,
$I=0,1,\ldots,n$ and $\eta_{IJ}={\rm diag}(-1,1,\ldots,1)$.
There are $n_V=n$ vector multiplets and thus $n$ complex scalars. If we choose
$Z^0=1$ and $Z^i=\tau^i$, $i=1,\ldots,n$, the symplectic vector becomes
\eq
v=\left(1,\tau^1,\ldots,\tau^n,-\frac i2,\frac i2\tau^1,\ldots,\frac i2\tau^n
\right)^T\,.
\feq
The K\"ahler potential and metric are
\eqn
&& e^{-{\cal K}}=1-|\vec\tau |^2, \\
&& g_{i\bar j}=\frac{\delta_{ij}}{1-|\vec\tau |^2}+\frac{\bar\tau^i\tau^j}
{(1-|\vec\tau |^2)^2}\,,
\feqn
where $|\vec\tau |^2=\sum_{i=1}^n\bar\tau^i\tau^i$. Positivity is ensured by
$|\vec\tau |^2<1$. In particular
\eq
X^0=\frac1{\sqrt{1-|\vec\tau |^2}}\,, \qquad\ X^i=\frac{\tau^i}{\sqrt{1-|\vec
\tau |^2}}
\feq
satisfy $\bar X^I\eta_{IJ}X^J=-1$. Defining $X_I=\eta_{IJ}X^J$, the kinetic
matrix for the vectors is
\eq
{\cal N}_{IJ}=-\frac i2 \eta_{IJ} - i X_I X_J\,.
\feq
Let us now look for a solution with
\eq
\tau^i=\tau^i(z)\,, \qquad\ b=iN(z)\,,
\feq
with $N(z)$ and $\tau^i(z)$ all real and positive. Then $\eta_{IJ}X^IX^J=-1$ so
that we get
\eq
(\text{Im}\,{\cal N})^{-1|IJ} = -2(\eta^{IJ}+2X^I X^J)\,,
\feq
and the scalar potential reads
\eq
V = 8g^2 - 16\frac{(g_0+\vec g\cdot\vec\tau)^2}{1-\vec\tau^{\,2}}\,,
\label{pot-XetaX}
\feq
where $g^2=g^Ig_I$, $g^I=\eta^{IJ}g_J$ and $\vec g\cdot\vec\tau=g_i\tau^i$.
\eqref{pot-XetaX} has an extremum for $\tau^i=-g_i/g_0$, with
$V|_{\text{extr.}}=24g^2$. In order to have a supersymmetric AdS vacuum, this
must be negative, so that $g_I$ is timelike, $g^2<0$.\\
Since
\eq
\text{Re}\,{\cal N} = 0\,, \qquad \frac{X^I}{\bar b}+\frac{\bar X^I}b = 0\,,
\qquad \frac{F_I}{\bar b}-\frac{\bar F_I}b = 0\,, 
\feq
\eqref{stab1} is satisfied for $q_I=0$. Moreover,
$\langle{\cal I}\,,d{\cal I}\rangle = 0$, hence (\ref{dsigma'}) holds as well.
If we define $H^I\equiv X^I/N$, \eqref{stab2} boils down to
\eq
e^{2\psi}[\partial_z H^I + 2g_J(-H^2\eta^{IJ} + 2H^IH^J)] = -2\pi p^I,
\label{stab2-XetaX}
\feq
with $H^2=\eta_{IJ}H^IH^J=-1/N^2$. Making use of the ansatz
\eq
\psi=\ln(az^2+c)\,, \qquad H^I=\frac{\alpha^I z+\beta^I}{az^2+c}
\feq
in eqns.~\eqref{stab2-XetaX} and \eqref{dzpsi}, one obtains the set of relations
\eq
a=2g_I\alpha^I\,, \qquad \alpha^I=\frac{2g^I\alpha^2}a\,, \qquad \alpha^Ic
- 2g^I\beta^2 = -2\pi p^I\,, \qquad g_I\beta^I=0\,, \label{relations}
\feq
where $\alpha^2=\eta_{IJ}\alpha^I\alpha^J$ and similar for $\beta^2$. Thus
\eq
N^2 =-\frac {(az^2+c)^2}{\alpha^2z^2 +\beta^2}\,, \label{lapse-XetaX}
\feq
which to be positive for large $z$ requires $\alpha^2<0$. This is indeed
satisfied, since $\alpha^I$ is proportional to $g^I$, and the latter is
timelike. The spacetime metric is
\eq
ds^2=-4N^2dt^2 + \frac{dz^2}{N^2} + (-\alpha^2 z-\beta^2)\,
e^{2\gamma}dw d\bar w\,,
\feq
while the scalar fields and fluxes read
\eq
\tau^i =\frac{\alpha^i z + \beta^i}{\alpha^0 z + \beta^0}\,, \qquad
F^I= 2\pi ip^Ie^{2\gamma}dw\wedge d\bar w\,,
\feq
so that the magnetic charges are $P^I=p^IV$, with $V$ given in \eqref{vol}.
Note that asymptotically for $z\to\infty$, we have $\tau^i\to -g_i/g_0$,
where the scalar potential becomes extremal. The spacetime approaches
AdS$_4$ in this limit. We assume $a>0$ and $c<0$ so that there is an event
horizon at $z_h=\sqrt{-c/a}$. In what follows, we shall again use the scaling
symmetries \eqref{scale1}, \eqref{scale2} to set $a=1$ and $\kappa=0,\pm 1$
without loss of generality. This implies then $\alpha^I=g^I/(2g^2)$.

It is easy to show that the positivity condition $\vec\tau^{\,2}<1$ for the
kinetic terms of the scalars is equivalent to $\alpha^2z^2+\beta^2<0$. From
\eqref{lapse-XetaX} one sees that this coincides with the condition of having a
positive lapse function in the region outside the horizon. If $\beta^2<0$,
this is always satisfied. Using
\eq
\frac\kappa4 = \frac c2 - 2g^2\beta^2\,, \label{kappa}
\feq
that follows from the third equation of \eqref{relations} by contracting with
$g_I$, one sees that $\kappa<0$ in this case. When $\beta^2\geq 0$ we have a
singularity at $z_s=(-\beta^2/\alpha^2)^{1/2}$, where $N^2$ diverges and
$\vec\tau^{\,2}=1$. Requiring this singularity to be hidden by the horizon
($z_h>z_s$) leads to $-c>-4g^2\beta^2$. Plugging this into \eqref{kappa} yields
again $\kappa<0$. Thus the geometry of the horizon is always hyperbolic.
Finally, taking into account that \eqref{relations} imply $p^I=g^I/(8\pi g^2)$,
we get for the entropy density
\eq
\frac SV = -2\pi^2p^2\,,
\feq
which depends on the charges only (and is positive since
$p^2=\eta_{IJ}p^Ip^J<0$).

\section{General near-horizon analysis}
\label{gen-nh}

We now want to analyze the near-horizon limit of a general static supersymmetric black hole solution
of the theory under consideration. This will be done without specifying a prepotential, with the aim
to obtain the analogue of the attractor equations in gauged supergravity.

As we are interested in the near horizon limit, the scalar fields are taken to
be constant. In order to get a spacetime geometry of the product form
AdS$_2$ $\times$ $\Sigma$, where $\Sigma$ denotes a two-dimensional space with
constant curvature, i.e., S$^2$, $\bR^2$, H$^2$ or a compact quotient thereof,
we must have $|b|^{-1}e^{\psi}=c$, with $c$ an arbitrary positive constant.
Using this, \eqref{dsigma'} and \eqref{dzpsi} can be easily integrated, with
the result
\eq
b = 4ig_I\bar X^Iz + b_0\,,
\feq
where $b_0$ denotes a complex integration constant. \eqref{stab1} and
\eqref{stab2} lead respectively to
\begin{eqnarray}
q_I &=& \frac{c^2}{2\pi}\left[4\text{Re}(F_Ig_J\bar X^J) + g_J\text{Re}\,
{\cal N}_{IL}(\text{Im}\,{\cal N})^{-1|JL}\right]\,, \\
p^I &=& \frac{c^2}{2\pi}\left[4\text{Re}(X^Ig_J\bar X^J) + g_J(\text{Im}\,
{\cal N})^{-1|IJ}\right]\,.
\end{eqnarray}
Note that this can be written more compactly as
\eq
\left(\begin{array}{c} p^I \\ q_I \end{array}\right) = \frac{c^2}{2\pi}
\left[4\text{Re}({\cal V}g_J\bar X^J) + {\cal M}\Omega{\cal G}\right]\,,
\label{charge-vec}
\feq
where $\cal M$ is the matrix introduced in \cite{Ceresole:1995ca},
\eq
{\cal M}Ê= \left(\begin{array}{cc} (\text{Im}\,{\cal N})^{-1} & (\text{Im}\,{\cal N})^{-1}\text{Re}\,{\cal N} \\
\text{Re}\,{\cal N}(\text{Im}\,{\cal N})^{-1} & \text{Im}\,{\cal N}+\text{Re}\,{\cal N}(\text{Im}\,{\cal N})^{-1}
\text{Re}\,{\cal N}\end{array}\right)\,,
\feq
$\Omega$ denotes the symplectic metric,
\eq
\Omega = \left(\begin{array}{cc} 0 & 1 \\ -1 & 0\end{array}\right)\,,
\feq
and we defined ${\cal G}=(g_J,0)^T$.

Then the fluxes are given by
\eq
F^I = -16\,\text{Im}(X^Ig_J\bar X^J)\,dt\wedge dz + 2\pi i p^I e^{2\gamma}dw
\wedge d\bar w\,.
\feq
The magnetic and electric charges read respectively
\eq
P^I = \frac1{4\pi}\int_{\Sigma_{\infty}}F^I = p^IV\,, \qquad
Q_I = \frac1{4\pi}\int_{\Sigma_{\infty}}G_I = q_IV\,,
\feq
where $G_{+I}={\cal N}_{IJ}F^{+J}$ \cite{Vambroes} and $V$ is given in
\eqref{vol}.
By using some relations of special geometry, one obtains for the central charge
\eq
Z = X^IQ_I - F_IP^I = \frac{Vc^2}{2\pi}ig_JX^J\,. \label{Z}
\feq
Before we continue, a small digression on the constant $c$ introduced above
is in order.
From \eqref{dzpsi} it is clear that $\psi$ is defined only up to an
arbitrary constant, so that we are free to shift
\eq
\psi \to \psi - \ln\lambda\,, \label{shift-psi}
\feq
which implies $c\to c/\lambda$, so that one can set $c$ equal to any value.
In order for $\Phi$ to be invariant, \eqref{shift-psi} must be
compensated by a shift in $\gamma$,
\eq
\gamma \to \gamma + \ln\lambda\,. \label{shift-gamma}
\feq
From the Liouville equation \eqref{liouville} we see that the magnetic
charge densities scale as $p^I\to p^I/\lambda^2$. Using \eqref{shift-gamma} in
\eqref{vol}, one gets $V\to \lambda^2V$, so that the product $c^2V$, and thus
the charges $P^I$, $Q_I$ and $Z$, remain invariant, as it must be.

Finally, the entropy of the black hole with this near-horizon geometry is
\eq
S = \frac{A_{\text{hor}}}4 = \frac{c^2V}4\,.
\feq
Using the expression \eqref{Z} for the central charge, this can be rewritten
as
\eq
S = \frac{\pi Z}{2ig_JX^J}\,. \label{S}
\feq
Multiplying \eqref{charge-vec} with $V$ and eliminating $c^2V$ by means of
\eqref{Z}, one gets
\eq
\left(\begin{array}{c} P^I \\ Q_I \end{array}\right) = \frac Z{ig_JX^J}
\left[4\,\text{Re}({\cal V}g_K\bar X^K) + {\cal M}\Omega{\cal G}\right]\,,
\label{attractor}
\feq
which represents the analogue of the attractor
equations \cite{Ferrara:1996dd,Ferrara:1996um}
for static supersymmetric black holes in gauged supergravity. If one were
able to solve \eqref{attractor} in order to obtain the moduli in terms
of the charges, one could plug the result into \eqref{S} to show that the
entropy does not depend on the values of the scalars on the horizon.
However, the eqns.~\eqref{attractor} are nonlinear, and in general might
not be invertible. In fact, in section \ref{X0X1} we encountered an explicit
example where invertibility breaks down: The horizon value of the scalar
$\tau$ is given in terms of the arbitrary (charge-independent) constant $\nu$,
i.~e., $\tau$ is not stabilized; in other words the black hole potential
has a flat direction. Nevertheless, as can be seen from \eqref{entr-X0X1},
the entropy is completely determined by the charges\footnote{Note that in
section \ref{X0X1} we have scaled $\kappa$ to be $-1$, which implies that the
volume $V$ is independent of the moduli, so that $S$ depends on the product
$P^0P^1$ only.}, and is thus still independent of $\nu$, in agreement with
the attractor mechanism.
Unfortunately we do not know of any way to show this for the case
\eqref{S} of a generic prepotential.

Let us take a closer look at the SU(1,1)/U(1) model of section \ref{X0X1},
without making the assumption $\tau=\bar\tau$ that was adopted there in
order to obtain an explicit black hole solution. We wish to determine the
moduli space spanned by the flat directions in the black hole potential.
To this end, we write down the attractor equations for the parametrization
\eqref{v-X0X1}. It is easy to show that \eqref{attractor} are equivalent to
\eq
Q_0 = Q_1 = 0\,, \qquad g_0P^0 - g_1P^1 = 0\,,
\feq
and thus the attractor equations imply only some constraints on the charges,
but do not involve the complex scalar $\tau$, which remains completely
arbitrary. The moduli space of BPS attractors with prepotential $F=-iX^0X^1$
is therefore SU(1,1)/U(1). Notice that in ungauged ${\cal N}=2$, $D=4$
supergravity coupled to abelian vector multiplets, there are no flat directions
in the 1/2-BPS attractor flow \cite{Ferrara:1997tw} (at least as long as the
metric of the scalar manifold is strictly positive definite),
but there is a nontrivial moduli space for non-BPS
flows \cite{Ferrara:2007tu,Bellucci:2008sv}. The new feature appearing in
gauged supergravity is thus the presence of flat directions in the black
hole potential also in the BPS case.

\acknowledgments

This work was partially supported by INFN and MIUR-PRIN contract 20075ATT78.
We would like to thank Bianca Letizia Cerchiai, Alessio Marrani, Andrea Mauri
and Wafic A.~Sabra for useful discussions, and Andrea Borghese and Diego S.~Mansi
for collaboration in the early stages of this project.

\end{document}